\documentclass{elsart}

\usepackage{graphicx}
\usepackage{amssymb}

\def\ltsim{\footnotesize{\mathop{\raisebox{-.4ex}{\rlap{$\sim$}}
\raisebox{.4ex}{$<$}}}}

\renewcommand{\theta}{\vartheta}

\renewcommand{\epsilon}{\varepsilon}

\begin{document}

%\tableofcontents
%\newpage

\begin{frontmatter}

% Title, authors and addresses

% use the thanksref command within \title, \author or \address for footnotes;
% use the corauthref command within \author for corresponding author footnotes;
% use the ead command for the email address,
% and the form \ead[url] for the home page:
% \title{Title\thanksref{label1}}
% \thanks[label1]{}
% \author{Name\corauthref{cor1}\thanksref{label2}}
% \ead{email address}
% \ead[url]{home page}
% \thanks[label2]{}http://nemoweb.lns.infn.it/
% \corauth[cor1]{}
% \address{Address\thanksref{label3}}
% \thanks[label3]{}

% \title{Possible detection of TeV neutrinos from Galactic Microquasars with an
% underwater \v{C}erenkov km${\bf ^3}$ telescope}

\title{Measurement of the atmospheric muon flux with the NEMO Phase-1 detector}

\author[INFNCT]{S. Aiello},
\author[INFNRM]{F. Ameli},
\author[INFNLNS,UniCT]{I. Amore},
\author[INFNGE]{M. Anghinolfi},
\author[INFNLNS]{A. Anzalone},
\author[INFNNA,UniNA]{G. Barbarino},
\author[INFNGE]{M. Battaglieri},
% \author[INFNBA,UniBA]{R. Bellotti},
\author[INFNBO,UniBO]{M. Bazzotti},
\author[INFNGE,UniGE]{A. Bersani},
\author[INFNPI,UniPI]{N. Beverini},
\author[INFNBO,UniBO]{S. Biagi},
\author[INFNRM,UniRM]{M. Bonori},
\author[INFNPI,UniPI]{B. Bouhadef},
\author[UniGE]{M. Brunoldi},
\author[INFNLNS]{G. Cacopardo},
\author[INFNRM,UniRM]{A. Capone},
\author[INFNCT]{L. Caponetto\thanksref{IPNL}},
\author[INFNBO,UniBO]{G. Carminati},
% \author[INFNBA]{B. Cassano},
%\author[INFNPI,UniPI]{E. Castorina},
%\author[INFNBA]{A. Ceres},
\author[INFNBO,UniBO]{T. Chiarusi},
\author[INFNBA]{M. Circella},
\author[INFNLNS]{R. Cocimano},
\author[INFNLNS]{R. Coniglione},
\author[INFNLNF]{M. Cordelli},
\author[INFNLNS]{M. Costa},
\author[INFNLNS]{A. D'Amico},
\author[INFNPI,UniPI]{G. De Bonis},
\author[INFNBA,UniBA]{C. De Marzo\thanksref{Dead}},
\author[INFNNA]{G. De Rosa},
\author[INFNBA]{G. De Ruvo},
\author[INFNGE]{R. De Vita},
\author[INFNLNS]{C. Distefano\corauthref{ca:fax}}\ead{distefano\_c@lns.infn.it},
\author[INFNPI,UniPI]{E. Falchini},
\author[INFNPI,UniPI]{V. Flaminio},
\author[INFNGE]{K. Fratini},
\author[INFNBO,UniBO]{A. Gabrielli},
\author[INFNLNS,UniCT]{S. Galat\`a\thanksref{CPPM}},
\author[INFNBO,UniBO]{E. Gandolfi},
\author[INFNBO,UniBO]{G. Giacomelli},
\author[INFNBO,UniBO]{F. Giorgi},
\author[INFNRM,UniRM]{G. Giovanetti},
\author[INFNCT]{A. Grimaldi},
\author[INFNLNF]{R. Habel},
\author[INFNLNS]{M. Imbesi},
\author[INFNGE]{V. Kulikovsky},
\author[INFNLNS,UniCT]{D. Lattuada},
\author[INFNCT,UniCT]{E. Leonora},
\author[INFNRM]{A. Lonardo},
% \author[INFNNA,UniNA]{G. Longo},
\author[INFNCT,UniCT]{D. Lo Presti},
\author[INFNRM,UniRM]{F. Lucarelli},
\author[INFNPI,UniPI]{A. Marinelli},
\author[INFNBO,UniBO]{A. Margiotta},
\author[INFNLNF]{A. Martini},
\author[INFNRM,UniRM]{R. Masullo},
% \author[INFNBA,UniBA]{R. Megna},
\author[INFNLNS,UniCT]{E. Migneco},
\author[INFNGE]{S. Minutoli},
% \author[INFNBA]{M. Mongelli},
\author[INFNPI,UniPI]{M. Morganti},
\author[INFNGE]{P. Musico},
\author[INFNLNS]{M. Musumeci},
\author[INFNRM]{C.A. Nicolau},
\author[INFNLNS]{A. Orlando},
\author[INFNGE]{M. Osipenko},
% \author[INFNNA]{G. Osteria},
\author[INFNLNS]{R. Papaleo},
\author[INFNLNS]{V. Pappalardo},
%\author[INFNCT,UniCT]{C. Petta},
\author[INFNLNS]{P. Piattelli},
\author[INFNGE]{D. Piombo},
\author[INFNLNS]{G. Raia},
\author[INFNCT]{N. Randazzo},
\author[INFNCT]{S. Reito},
\author[INFNGE,UniGE]{G. Ricco},
\author[INFNLNS]{G. Riccobene},
\author[INFNGE]{M. Ripani},
% \author[INFNLNS,UniCT]{D.J. Romeo\thanksref{nomore}},
\author[INFNLNS]{A. Rovelli},
\author[INFNBA,UniBA]{M. Ruppi},
\author[INFNCT,UniCT]{G.V. Russo},
\author[INFNNA,UniNA]{S. Russo},
\author[INFNLNS]{P. Sapienza},
\author[INFNCT]{D. Sciliberto},
\author[INFNLNS]{M. Sedita},
\author[MSU]{E. Shirokov},
\author[INFNRM,UniRM]{F. Simeone},
\author[INFNCT,UniCT]{V. Sipala},
\author[INFNBO,UniBO]{M. Spurio},
\author[INFNGE,UniGE]{M. Taiuti},
%\author[INFNPI]{G. Terreni},
\author[INFNLNF]{L. Trasatti},
\author[INFNCT]{S. Urso},
%\author[INFNLNF]{V. Valente},
\author[INFNRM,UniRM]{M. Vecchi},
\author[INFNRM]{P. Vicini},
\author[INFNRM]{R. Wischnewski}.

%%%%%%%%\thanks[X]{This is the history of the paper, etc etc}

\corauth[ca:fax]{Fax: +39 095 542 398}
\thanks[Dead]{Deceased}
\thanks[CPPM]{Present address, Centre de Physique des Particules de Marseille, CNRS/IN2P3 et Univ. de la M\'editerran\'ee, 163 Av. de Luminy, Case 902, 13288 Marseille Cedex 9, France}
\thanks[IPNL] {Present address, CNRS/IN2P3/IPNL, Domaine scientifique de la Doua, B$\hat{\hbox{a}}$timent Paul Dirac 4, Rue Enrico Fermi, Lyon, France}

%\thanks[UniLe]{Present address, Dipartimento di Fisica Universit\`a del Salento, Via Arnesano, 73100, Lecce, Italy }
%\thanks[UPV]{Present address, Universidad Politecnica de Valencia, Campus de Gandia, Ctra. Nazaret-Oliva, 46730, Grao de Gandia, Valencia, Spain}
%\thanks[UniCTDMFCI]{Present address, Dipartimento di Metodologie Fisiche e Chimiche per l'Ingegneria Universit\`a di Catania, Viale A. Doria 6, 95125 Catania, Italy }
%\thanks[nomore]{Presently in private occupation}

\address[INFNLNS]{Laboratori Nazionali del Sud INFN, Via S.Sofia 62, 95123, Catania, Italy}
\address[INFNLNF]{Laboratori Nazionali di Frascati INFN, Via Enrico Fermi 40, 00044, Frascati (RM), Italy}
\address[INFNBA]{INFN Sezione Bari, Via Amendola 173, 70126, Bari, Italy}
\address[INFNBO]{INFN Sezione Bologna, V.le Berti Pichat 6-2, 40127, Bologna, Italy}
\address[INFNCT]{INFN Sezione Catania, Via S.Sofia 64, 95123, Catania, Italy}
\address[INFNGE]{INFN Sezione Genova, Via Dodecaneso 33, 16146, Genova, Italy}
\address[INFNNA]{INFN Sezione Napoli, Via Cintia, 80126, Napoli, Italy}
\address[INFNPI]{INFN Sezione Pisa, Polo Fibonacci, Largo Bruno Pontecorvo 3, 56127, Pisa, Italy}
\address[INFNRM]{INFN Sezione Roma 1, P.le A. Moro 2, 00185, Roma, Italy}

\address[UniBA]{Dipartimento Interateneo di Fisica Universit\`a di Bari, Via Amendola 173, 70126, Bari, Italy}
\address[UniBO]{Dipartimento di Fisica Universit\`a di Bologna, V.le Berti Pichat 6-2, 40127, Bologna, Italy}
\address[UniCT]{Dipartimento di Fisica e Astronomia Universit\`a di Catania, Via S.Sofia 64, 95123, Catania, Italy}
\address[UniGE]{Dipartimento di Fisica Universit\`a di Genova, Via Dodecaneso 33, 16146, Genova, Italy}
\address[UniNA]{Dipartimento di Scienze Fisiche Universit\`a di Napoli, Via Cintia, 80126, Napoli, Italy}
\address[UniPI]{Dipartimento di Fisica Universit\`a di Pisa, Polo Fibonacci, Largo Bruno Pontecorvo 3, 56127, Pisa, Italy}
\address[UniPV]{Centro Interdisciplinare di Bioacustica e Ricerche Ambientali, Dipartimento di Biologia Animale Universit\`a di Pavia, Via Taramelli 24, 27100, Pavia, Italy}
\address[UniRM]{Dipartimento di Fisica Universit\`a ``Sapienza'', P.le A. Moro 2, 00185, Roma, Italy}
\address[MSU]{Faculty of Physics, Moscow State University, 119992, Moscow, Russia}

\begin{abstract}
% Text of abstract
The NEMO Collaboration installed and operated an underwater
detector including prototypes of the critical elements of a
possible underwater km$^3$ neutrino telescope: a four-floor tower
(called Mini-Tower) and a Junction Box. The detector was developed
to test some of the main systems of the km$^3$ detector, including
the data transmission, the power distribution, the timing
calibration and the acoustic positioning systems as well as to
verify the capabilities of a single tridimensional detection
structure to reconstruct muon tracks. We present results of the
analysis of the data collected with the NEMO Mini-Tower. The
position of photomultiplier tubes (PMTs) is determined through the
acoustic position system. Signals detected with PMTs are used to
reconstruct the tracks of atmospheric muons. The angular
distribution of atmospheric muons was measured and results
compared to Monte Carlo simulations.
\end{abstract}

\begin{keyword}
% keywords here, in the form: keyword \sep keyword
Atmospheric muons \sep Neutrino telescopes \sep NEMO
% PACS codes here, in the form: \PACS code \sep code
\PACS 95.55.Vj \sep %Neutrino, muon, pion, and other elementary particle detectors; cosmic ray detectors
95.85.Ry \sep %Neutrino, muon, pion, and other elementary particles; cosmic rays
96.40.Tv  %Neutrinos and muons
\end{keyword}
\end{frontmatter}

\section{Introduction}
\label{sec:introduction}

High energy neutrinos are considered optimal probes to identify
the sources of high energy cosmic rays. Many indications suggest
that cosmic objects, where acceleration of charged
particles takes place, should also produce high energy gamma ray
and neutrino fluxes. Indeed, $p \gamma$ or $p p$ interactions
responsible for $>$ TeV neutrinos and $\gamma$ rays, are
expected to occur in several astrophysical environments such as
supernova remnants, microquasars, gamma-ray bursts and active
galactic nuclei \cite{ref:Aharonian_Science_2007}. For many
sources the neutrino flux,  produced in the decay chains due to charge pions, is
expected to be similar to the flux of high-energy gamma rays of hadronic origin, which can be measured by TeV gamma ray telescopes such as MAGIC
\cite{ref:MAGIC_2005}, HESS \cite{ref:HESS_SNR_Nature2004} and
VERITAS \cite{ref:VERITAS_2008}. Because of the low expected
neutrino fluxes from galactic and extragalactic sources
\cite{modelli}, the effective opening of the high energy neutrino
astronomy era can really be made with detectors of km$^3$ scale.
After the success of the first  generation of underwater/ice
neutrino telescopes, such as BAIKAL \cite{baikal} and AMANDA
\cite{amanda}, the construction of the first km$^3$ telescope,
IceCube, \cite{icecube} started at the South Pole. The detector is announced to be completed by 2011. In the Mediterranean Sea, the
ANTARES telescope is taking data since 2006 in a partial
configuration and since 2008 in its full set-up \cite{antares}.
The ANTARES Collaboration together with the NESTOR \cite{nestor}
and NEMO \cite{nemo} Collaborations are conducting an intense R\&D
activity for the future km$^3$ Mediterranean telescope. Recently
the three collaborations joined their efforts to design the KM3NeT
undersea infrastructure that will host a km$^3$ telescope; its
construction is expected to start by 2013 \cite{km3net}.

The activity of the NEMO Collaboration was mainly focused on the
search and characterization of an optimal site for the detector
installation and on the development of key technologies for the
km$^3$ underwater telescope to be installed in the  Mediterranean
Sea. A deep sea site with optimal features in terms of depth and
water optical properties was identified at a depth of 3500 m about
80 km off-shore from Capo Passero (Southern cape of Sicily).  A
long term monitoring of the site has been carried out \cite{sito}. 

An other effort of the NEMO Collaboration was the
definition of a feasibility study of the km$^3$ detector, which
included the analysis of the construction and installation
issues and the optimization of the detector geometry by means of
numerical simulations. 

To ensure an adequate process of
validation of the proposed solutions, a technological demonstrator was built and installed
off-shore the port of  Catania (Sicily, Italy). This project, called NEMO
Phase-1, allowed the test and the qualification of the key
technological elements (mechanics, electronics, data transmission,
power distribution, acoustic positioning and time calibration
system) proposed for the km$^3$ detector  \cite{Migneco08}.
Some of the technical solutions, further developed within the KM3NeT Consortium, have been proposed
for the km$^3$ detector construction.

In this paper, after a detailed description of the NEMO Phase-1
lay-out and operation, we focus on the atmospheric muon data
analysis procedure and present the main results. In particular, the
atmospheric muon angular distribution was measured and compared
with Monte Carlo simulations. The vertical muon flux was
determined from the angular distribution, computing the related
depth intensity relation. Results were compared with theoretical
predictions and with results of other experiments.

\section{The NEMO Phase-1 detector}
\label{sec:operations}

\begin{figure}[htb]
\begin{center}
\includegraphics[height =8cm]{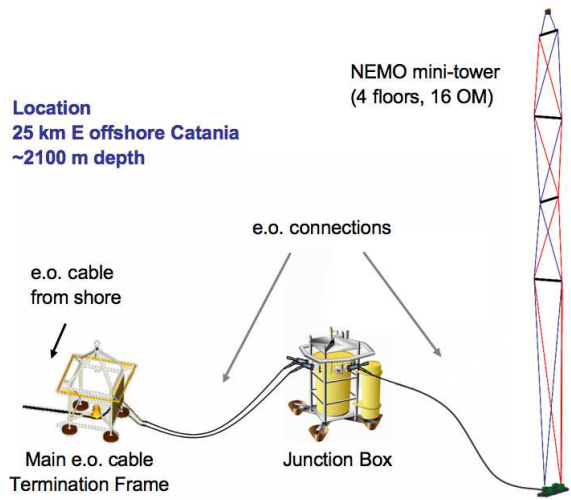}
\end{center}
\caption{Lay-out of the NEMO Phase-1 installation at the Catania
Test Site.} \label{fig:ts}
\vspace*{1cm}
\end{figure}

\begin{figure}[h]
\begin{center}
\includegraphics[height =7.5cm]{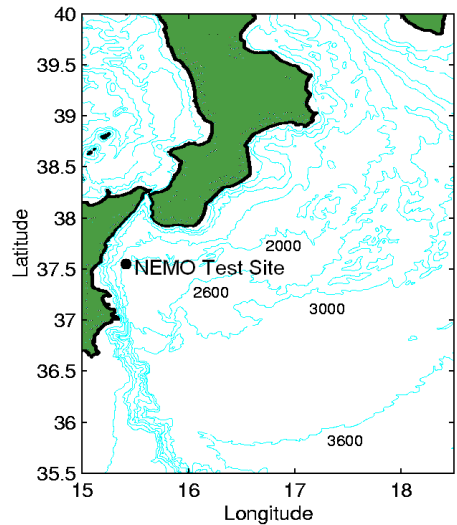}
\end{center}
\caption{Location of the Underwater Test Site of the Laboratori Nazionali del Sud.}
\label{fig:testsite}
\end{figure}

The NEMO Phase-1 apparatus is composed of two main elements: the  Mini-Tower and
the Junction Box (JB) interconnected as sketched in
Fig. \ref{fig:ts} and described in the following.
NEMO Phase-1 was installed between 10 and 19
December 2006 at the underwater Test Site of the Laboratori
Nazionali del Sud, off shore Catania at a depth of 2080 m,
latitude: 37$^\circ$ 33' 4" N and longitude: 15$^\circ$ 23' 2'' E
(Fig. \ref{fig:testsite}).
The operation was conducted with the cable layer vessel TELIRI
\cite{Elettra}. The JB and the tower were deployed from the sea
surface by means of a winch and positioned on the seabed with an
accuracy of a few meters.

The JB was then connected to the main cable
termination and the tower to the JB with electro-optical links
equipped with wet mateable hybrid connectors. The connection
operations were performed with an underwater Remotely Operated
Vehicle (ROV). The operation was completed with the successful
unfurling of the tower that assumed the correct configuration. All
active elements, such as PMTs, electronics, acoustic positioning,
data transmission and acquisition, worked correctly.

After four weeks of successful operation, two problems were
encountered, which prevented a fully efficient exploitation of the
apparatus. Firstly, a loss of buoyancy of the main buoy of the Mini-Tower occurred, which caused a slow sinking of the whole tower; this problem was later ascribed to a poor manufacturing process of the buoy. Then, about four months
after the connection, the optical
fiber transmission started to show a significant attenuation at the JB level. The JB was
recovered in June 2007. After tests in a hyperbaric chamber, the
problem was identified in a faulty optical penetrator. The problem was solved by bypassing the connector with a redesign of the optical system
lay-out. The JB was re-installed in April 2008 and proved to work correctly.

In spite of these problems, the NEMO Phase-1 allowed to validate a number of
technical solutions for the km$^3$ telescope and to demonstrate
the capability of the NEMO tower to detect and track muons. During
5 months of operation about 500 GB of data were recorded.

\subsection{The Junction Box}

The JB provided connection between the main electro-optical cable
and the detector structures and was designed to host and protect
from the effects of corrosion and pressure the opto-electronic
boards dedicated to the distribution and the control of the power
supply and digitized signals.

The NEMO Phase-1 JB was built following the concept of double
containment. Pressure resistant steel vessels were hosted inside a
large fiberglass container, filled with silicon oil and
pressure compensated. This solution has the advantage to decouple
the two problems of pressure and corrosion resistance. All the electronics components that were proven able to withstand
high pressure were installed directly in the oil bath
\cite{Migneco08}. The JB was equipped with 6 (2 inputs and 4
outputs) hybrid electro-optical wet mateable bulkheads. The two
inputs were connected, by means of two underwater cables, to the two outputs of the termination frame of the main electro-optical cable, while one of the four available outputs of the JB was used to connect the Mini-Tower by means of a 300 m electro-optical link. The JB was equipped with a HV transformer and switches to distribute power to the output connectors, and with a system of optical couplers to connect the fibers from the input to the output connectors.

\subsection{The Mini-Tower}
\label{sec:minitower}

The  Mini-Tower was a small-size prototype of the NEMO Tower
\cite{Migneco06}. It was a three dimensional flexible structure
composed by a sequence of four horizontal elements ({\it floors})
interlinked by a system of tensioning ropes and anchored on the
seabed. The structure was kept vertical by an appropriate buoyancy
on the top.

\begin{figure}[h]
\begin{center}
\vspace{0.5 cm}
\includegraphics[height =10cm]{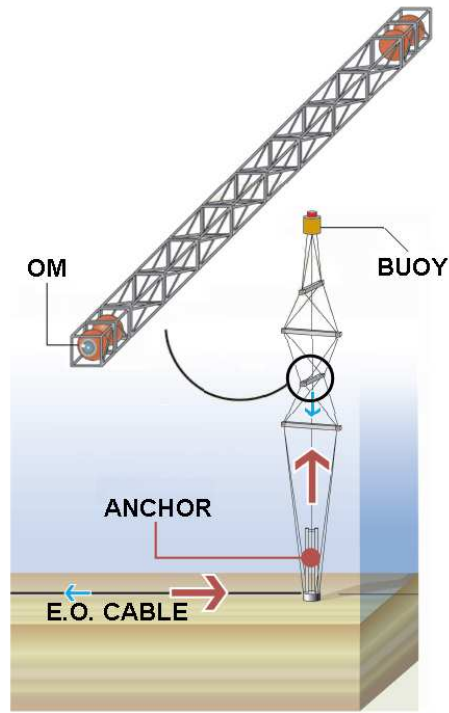}
\end{center}
\caption{Sketch of the NEMO Mini-Tower, in which 4 floors were equipped with a total of 16 PMTs (the electronics positioned
along the beam is not indicated, see Fig. \ref{fig:schema}).}
\label{fig:torre}
\end{figure}

The storey was a 15 m long structure supporting two Optical
Modules (OMs), one down-looking and one horizontally looking, at
each end: four OMs were installed on each storey. Each floor was
connected to the following one by means of four ropes arranged in
such a way to force each floor to a position perpendicular to its
vertical neighbors. The floors were vertically spaced by 40 m.  An
additional spacing of 100 m was present
between the tower base and the lowermost floor (Fig.
\ref{fig:torre}).

\begin{figure}[htb]
\begin{center}
\includegraphics[height =12cm]{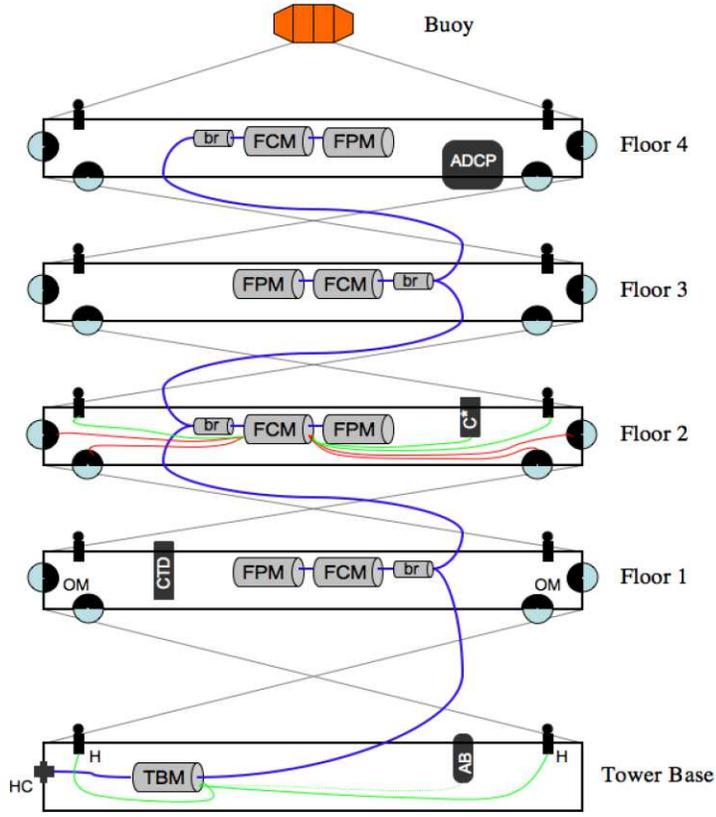}
\end{center}
\caption{Cabling layout of the NEMO Mini-Tower, including the
Tower Base Module (TBM);  for each of the 4 floors are indicated
the floor breakouts (br), the Floor Control Modules (FCM) and the
Floor Power Modules (FPM). Connection to the Junction Box is
provided through a wet mateable hybrid connector (HC) on the tower
base.} \label{fig:schema}
\end{figure}

In addition to the 16 OMs the instrumentation installed on the
Mini-Tower included several sensors for calibration and
environmental monitoring: an Acoustic Doppler Current Profiler (RD
Instruments Workhorse ADCP) to measure water current; a light
transmissometer (Wetlabs C*)  to measure water transparency; a
Conductivity--Tempe\-rature--Depth probe (Sea-Bird Electronics
37-SM microcat CTD) to monitor sea water properties; a pair of
hydrophones on each floor and on the tower base for acoustic
positioning. A scheme of the instrumentation is shown
in Fig. \ref{fig:schema}.

The NEMO tower was designed to be assembled in a compact
configuration, also kept during the transport and the deployment,
which was performed from a surface vessel by means of a winch.
After the positioning on the seabed and the connection to the
undersea cable network, the tower was unfurled thanks to the pull
provided by the buoy. This procedure was actuated remotely from
the surface vessel by means of an acoustic device.

%\section{The Mini-Tower subsystems}

\subsection{The Optical Module}
\label{sec:om}

Each optical module used for the NEMO Phase-1 detector was
composed by a PMT enclosed in a 17" pressure resistant sphere of
thick glass. The PMT was a 10" Hamamatsu R7081Sel with 10
stages \cite{Leonora09}. In spite of its large photocathode area, this PMT has a
good time resolution, about 3 ns FWHM for single photoelectron
pulses, with a charge resolution of 35\%. Mechanical and optical
contact between the PMT and the internal glass surface was ensured
by an optical silicon gel. A $\mu$-metal cage shielded the PMT
from the Earth magnetic field. The base card circuit for the high
voltage distribution (Iseg PHQ 7081SEL) required only a low
voltage supply (+5 VDC) and generated all necessary voltages for
cathode, grid and dynodes with a power consumption of less than
150 mW.

A Front-end Electronics Module
(FEM), built with discrete components \cite{Nicolau06}, was also
placed inside the OM. The FEM performed sampling at 200 MHz by means of
two 100 MHz staggered Flash ADCs, whose outputs were ``captured''
by a Field Programmable Gate Array (FPGA). The FPGA performed threshold discrimination, stored it with an event time stamp in an
internal 16 kbit FIFO, packed OM data and local slow control
information and coded everything into a bit stream frame ready to
be transmitted on a twisted pair at 20 Mb/s. The main
features of this solution are the moderate power consumption, the
high resolution and the large input dynamics range obtained by a
quasi-logarithmic analog compression circuit, with a raw time resolution of 5 ns. The FEM board digitized and sent
pulse waveform information up to a maximum pulse rate of  $\sim$150 kHz.
The FEM board was also equipped with analog and digital electronics
to control the OM power supply  and 
to monitor the operating conditions, such of temperature, humidity and electrical parameters. It could also perform a calibration of
the non linear response of the logarithmic compressor.

In addition, the board provided an estimate of the average count
rate by  counting the number of hits with amplitude exceeding a
threshold of 0.3 single photo-electron (s.p.e.) in a 10 ms time
window as shown in sec. \ref{sec:pmtrates}. This estimate did not
suffer from the limitation of the data transfer process and
allowed to measure the signal rate up to 6.5 MHz.

\subsection{The data acquisition system}
\label{sec:daq}

The design of the data acquisition system for NEMO Phase-1 was
based on technical choices that allow scalability to a much larger
apparatus \cite{Bunkheila06,Ameli08}.

The optical connection between the counting room facilities
on-shore and the  detector under water was driven by pairs of twin
electronic boards, called Floor Control Module (FCM), located at
both extremities  of each optical link, and deputed to manage
either the OM  data stream, from off-shore to on-shore, and  the Slow
Control commands, sent in the opposite direction. Each FCM board
was  powered by a Floor Power Module (FPM) lodged on the
corresponding floor, and was connected to the Tower Base Module
(TBM) through an optical fiber backbone. The TBM was connected
through an inter-link cable to the JB. A detailed description of
the Mini-Tower electronics is given in \cite{Ameli08}.

The OM data acquisition worked as follows: each off-shore FCM
multiplexed the signal  produced by the corresponding four OM-FEM
pairs in a floor, converted them from electric to optical and sent
it on-shore through the optical link. Each on-shore FCM, hosted on
a dedicated server, de-multiplexed the incoming data and
distributed it to the Trigger and Data Acquisition System, which
was composed of the MasterCPU server, for data filtering, and a
post-trigger data storage facility. Each on-shore server was
connected to the others via  a standard 1 Gigabit Ethernet
network.

In order to reconstruct muon tracks using the \v{C}erenkov
technique, a common timing must be known in the whole apparatus at
the level of each detection device to allow time correlation of
events. For this reason a synchronous and fixed latency protocol,
which embedded data and clock timing in the same serial bit
stream, was used for communications between onshore and offshore. The implemented system 
relied on Dense Wavelength
Division Multiplex (DWDM) techniques, using totally passive
components with the only exception of the line termination
devices, i.e. electro-optical transceivers. The precision of the
common timing of the apparatus was measured and monitored 
using a dedicated time calibration system, as illustrated in the next subsection.

\subsection{The time calibration system}

The time calibration was performed with an embedded system to track the possible
drifts of the time offsets during the operations of the apparatus
underwater \cite{Ruppi06}. This system measured the offsets with
which the local time counters inside the optical modules were
reset on reception of the reset commands broadcasted from shore.
The operation was performed with a redundant system: (1)
a two-step procedure for measuring the offsets in the time
measurements of the optical sensors and (2) an all-optical
procedure for measuring the differences in the time offsets of the
different optical modules.

In the first system the needed measurements were performed in two
separate steps: using an ``echo"  timing calibration and using an
``optical"  timing calibration. The former allowed the measurement
of the time delay for the signal propagation from the shore to the
FCM of each floor; the latter, based on a network of optical
fibres which distributes calibration signals from fast light
pulsers to the OMs, allowed to determine the time offsets between
the FCM and each optical module connected to it.

The second system was an extension of the
optical timing calibration system, which allowed to simultaneously
calibrate the optical modules of different floors of
the tower.

\subsection{The acoustic positioning system}

Another key requirement for the muon tracking with an underwater Cherenkov apparatus is the knowledge of each optical sensor
position. While the position and orientation of the tower base was
fixed and known from its installation, the rest of the structure
could bend under the influence of sea currents. A precise
determination of the position of each tower floor was achieved by
means of triangulations performed by measuring 
the propagation times of acoustic signals between a Long-Base-Line (LBL) of acoustic
beacons, placed on the sea floor, and a couple of hydrophones
(labeled H0 and H1) installed on each tower floor close to the
positions of the optical modules. The inclination and orientation
of each tower floor was also measured by a tiltmeter-compass board
placed inside the FCM.

\begin{figure}[h]
\begin{center}
\includegraphics[width=9.5cm]{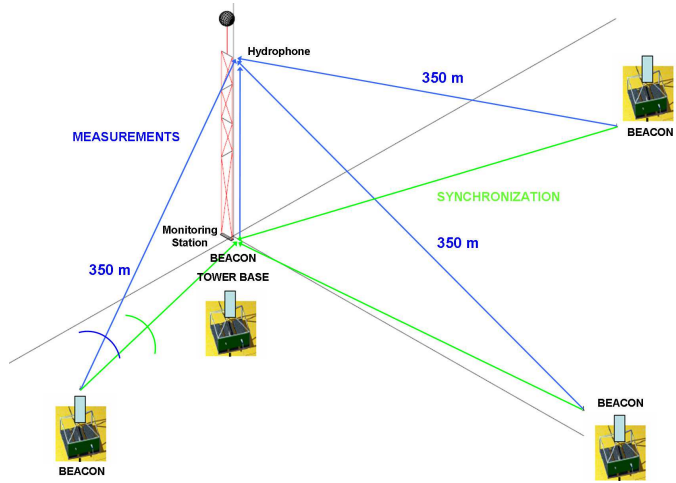}
\end{center}
\caption{Schematic view of the Acoustic Positioning System (APS)
and of the Long-Base-Line (LBL) configuration.} \label{fig:aco}
\end{figure}

The LBL was realized with four stand-alone battery-powered
acoustic beacons and one additional beacon located on the tower
base (Fig. \ref{fig:aco}).
To recognize the beacon pulses a technique called Time Spectral
Spread Codes (TSSC) was used. Each beacon transmitted 
a unique pattern of 6 pseudo-random pulses (spaced by 1 s). The duration of each pulse 
was 5 ms and the sequence of
pulses was built to avoid overlap between consecutive pulses.
In this way a typical beacon pulse sequence was recognized without
ambiguity and all the beacons could transmit their characteristic
pulse sequence at the same acoustic frequency. This was an
advantage since all beacons could be identical except for the
software configuration that defines the pulse sequence; the
receivers were then sensitive to the 32 kHz acoustic channel.
Since the beacon clocks were autonomous, in order to determine the position
of the hydrophones,  the LBL had to be synchronized to the master
clock of the apparatus. This synchronisation took place
acoustically using a monitoring station, formed by two
hydrophones, connected to the Mini-Tower data acquisition system,
placed in correspondence to the tower base. Prior to the
connection of the Mini-Tower, the four acoustic beacons providing
the LBL were deployed around the apparatus at an approximate
distance of 250 m from the tower base. In order to obtain the required accuracy of
$\ltsim$15 cm - comparable with the size of the PMT - the time of
flight of acoustic signals between the LBL beacons and the
monitoring station had to be evaluated with an accuracy of the
order of 10$^{-4}$ sec. To achieve this goal an accurate
calibration of the LBL was performed, taking into account the
clock drift of the stand-alone beacons. In particular, 
the absolute positions of the beacons and their relative distances
were determined,
acoustically, at the time of detector installation, using a ROV
equipped with a 32 kHz, GPS synchronized pinger and with a high accuracy pressure sensor.

%\section{Detector data analysis: control and environmental parameters}
\section{Control and environmental parameters}

All the Slow Control data (including data from all
environmental sensors and the acoustic positioning system)
were managed from shore by means of a dedicated Slow
Control Management System \cite{Rovelli06} and analyzed as described in the following.

\subsection{Acoustic positioning data}

The LBL calibration procedure allowed the
determination of distances between the beacons and the monitoring
station, and to calculate the time of flight of the acoustic pulses
as  the difference between the time of arrival of the
acoustic signal on each hydrophone and the time of emission of the
beacon. The sound velocity was calculated using the
CTD data. The time of emission of the beacon
pulse, in the common detector clock reference time, was obtained
measuring the time of arrival of this pulse at the monitoring
station. This procedure allowed to compensate for the clock drift of
the stand-alone beacons (about 20 ns/s) during the livetime of the
apparatus. In order to merge in post-processing the acoustic positions data together
with optical module detection information, both were time stamped
with a universal time reference tag.

The acoustic positioning system data were extensively analyzed. The tower positions
were reconstructed and the movements were measured as a function of time, on
long and short time scales \cite{Amore08}.

\begin{figure}[h]
\begin{center}
\includegraphics[width=10.5cm]{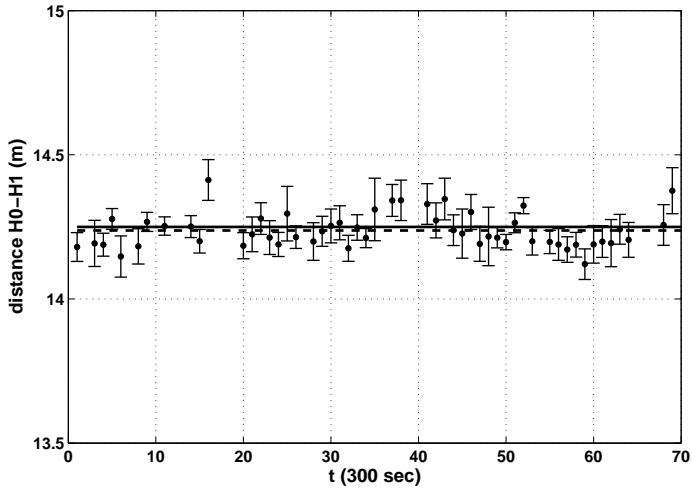}
\end{center}
\caption{Distance H0-H1 measured for floor 2. Each point is the
averaged distance over a period of 5 minutes, in the time interval
from 1$^{st}$ February h.17 to 1$^{st}$ February h.23 (6 hours).
The mean value of the measured distance is 14.24 $\pm$ 0.06 m
(dashed line). This value is compared with the construction
distance of 14.25 $\pm$ 0.01 m (solid line), measured on-shore,
during the tower integration.} \label{fig:accuracy}
\end{figure}

In order to estimate the accuracy of the positioning system, the distances between
hydrophones H0 and H1 on the same floor were
measured. In Fig.\ref{fig:accuracy} the distance H0-H1 measured
for floor 2 is shown. This result indicates that the obtained
accuracy in the determination of hydrophone positions is better
than 10 cm.

\subsection{PMT counting rates}
\label{sec:pmtrates}

The instantaneous  PMT rate values were computed by the FEMs as
described in sec. \ref{sec:om}. The PMT counting rates gave an
instantaneous estimate of the optical background during the
detector livetime.

\begin{figure}[h]
\begin{center}
\includegraphics[width=10.5cm]{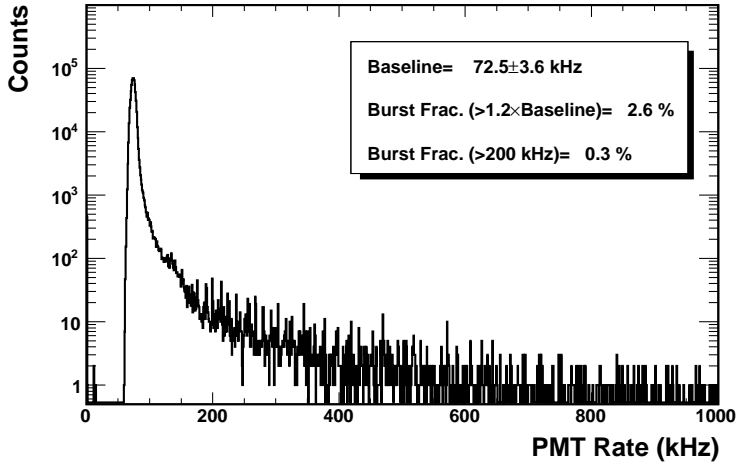}
\end{center}
\caption{Histogram of hit rate distribution from one PMT located
on the 4$^{\hbox{th}}$ floor, recorded in the period 10-20 January
2007.} \label{fig:rate}
\end{figure}

In Fig. \ref{fig:rate} the histogram of the rate
distribution is plotted for a PMT located on the 4$^{\hbox{th}}$
floor, in the time interval between 10 and 20 January 2007. The
histogram shows a peak in the 75-80 kHz interval due to $^{40}$K
decay plus a contribution due to diffuse bioluminescence
\cite{biolu}. The frequency value at the peak, commonly called the
\emph{baseline} of the optical background \cite{antaresnoise}, was
determined fitting the peak of the distribution with a Gaussian function. The
baseline obtained from the fit is 72.5$\pm$3.6 kHz. The
distribution, plotted in Fig. \ref{fig:rate}, shows also a tail
extending to several hundreds kHz due to intense bioluminescence
bursts. This contribution is measured by means of the so-called \emph{burst fraction} and it
was calculated, for comparison with the ANTARES detector data in
two different ways \cite{antaresnoise}:

\begin{itemize}
\item the percentage of time in which the rate exceeds 200 kHz;
\item the percentage of time in which the rate exceeds 1.2 times the baseline rate value.
\end{itemize}

For the data shown in Fig. \ref{fig:rate}, the percentage of time with rates exceeding 200 kHz was 0.3\% while the percentage of time with rates larger than 1.2 times the baseline rate was 2.6\%.

\begin{figure}[h]
\begin{center}
\vspace{0.8cm}
\includegraphics[width=11.5cm]{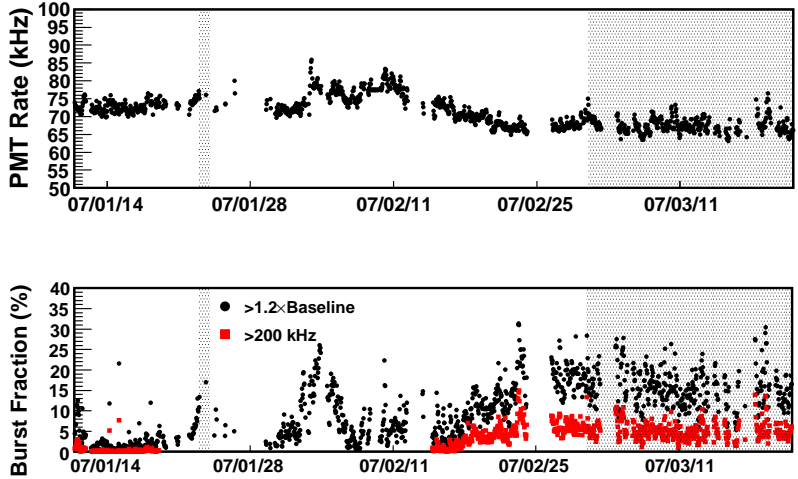}
\end{center}
\caption{Time dependence of the baseline rate (upper panel) and of
the burst fraction (lower panel) for a PMT located on the
4$^{\hbox{th}}$ floor in the period from 7 January (07/01/07) to
31 March 2007 (07/03/31). The shaded areas represent the two
periods of data acquisition when the muon trigger was active, see
text.} \label{fig:ratevstime}
\end{figure}

The baseline and burst fraction of the optical background were determined
for the full period of the detector operation. Typical results,
for an OM located in the 4$^{\hbox{th}}$ Mini-Tower floor, are
shown in Fig. \ref{fig:ratevstime}. The data indicate two time intervals with different
characteristics of the optical background. The first period, from
the detector activation to the first week of February 2007, was
characterized by a baseline rate of $\sim73$ kHz and a burst
fraction of the order of a few percents (with peaks up to 20\%)
while the second one, starting approximately on February 11,
characterized by a  slightly lower baseline rate (average $\sim67$
kHz) and a higher burst fraction ($\sim20\%$). These results can
be explained as due to two effects: a change of the water
properties and a change in the detector configuration due to the
loss of buoyancy of the tower. Starting from February 2007, in fact, the
detector fell down by about 40 meters; thus the 4$^{\hbox{th}}$
floor was at about 80 meters from the sea bottom. 
On the other hand during the same period of time
we observed an increase of the underwater currents \cite{Amore08}, and as 
shown by the ANTARES Collaboration \cite{antabiolum}, 
the level of bioluminescence is usually correlated with the value of the sea current. The somehow contradictory decrease of the baseline can be 
explained as a consequence of the increase of
water turbidity, due to particulate, close to the bottom and under
the effect of the increased current.

\section{Atmospheric muon analysis}

In order to reconstruct atmospheric muon tracks and to measure the
muon flux a dedicated data selection and analysis procedure was
applied. The atmospheric muon track
analysis was implemented for a limited period of the detector
livetime, shown by the shaded area in Fig. \ref{fig:ratevstime}:
from January 23 and 24 2007 -when the tower was completely
unfolded- and from March 2 to April 12 2007, when the two lowest
floors of the tower were laying on the seabed.

\subsection{The on-line muon trigger}

As described in section  \ref{sec:daq}, the data acquired by the OMs
were distributed, on-shore, to the MasterCPU.
On this machine, an on-line data selection algorithm was implemented to
select muon events among the optical background.

Since the main contribution to optical noise was due to
uncorrelated PMTs hits -with an average rate of about 70 kHz- the
selection  criterion (hereafter trigger seed) consisted in
imposing time coincidence in a narrow time-window of 20 ns among
hits  occurred on pairs of adjacent  PMTs (i.e. the two PMTs located
at the same  extremity of the same floor). This trigger seed is
called  {\em Simple Coincidence} (SC).

When an SC trigger occurred, the data  acquired by all the PMTs
within a {\em Triggered  Time Window} (TTW),  centered around the
SC time, were  stored on file and identified as possible muon
event for off-line data analysis.
If another trigger seed occurred in the TTW after the first one,
the TTW  itself was  extended by $\Delta t_{TTW}$ after the new
seed time.

In order to test the on-line selection algorithm, two different
lengths of the TTW were used, i.e. $\Delta t_{TTW} = \pm$2 $\mu$s
and  $\pm$5 $\mu$s, around the trigger seed time. It
was found that, after the application of the causality filter on
PMT hits (see sec. \ref{sec:causality}),  the two data sets  taken
with different $\Delta t_{TTW}$, were equivalent.

The measured total SC trigger rate ranged between 1.5 and 2 kHz,
while the expected  atmospheric muon signal  in the Mini-Tower,
evaluated  with Monte  Carlo  simulations, was $\sim$1 Hz  (see
sec.  \ref{sec:results}). Therefore, a further off-line data
selection, with more complex and more selective algorithms, too
slow for the requirements of the on-line trigger, was necessary.

\subsection{The off-line PMT data calibration}

The first step of the off-line event analysis was PMT data
calibration. PMT hits, recorded on each event file, were
decompressed and calibrated \cite{Simeone08}. First, the PMT hit
wave-form was re-sampled at 2 GHz (the ADC sampling is 200 MHz).
Therefore, the data values measured in ADC channels were decompressed 
and converted into
amplitudes (in mV unit), using a decompression table generated
during the FEM characterization phase. The 
rising edge of the waveform was then fitted with a sigmoid function and the hit time
was evaluated at the inflection point, and then corrected with the time offset provided etc.. At the end of this
process the PMT hit waveform was reconstructed: the integral
charge was determined with an uncertainty of $\sigma\sim0.3$ pC;
this value was converted in units of p.e. taking into account that
1 p.e. = 8 pC. The PMT hit calibration allowed to obtain
hit time evaluation with a precision of $\sigma\sim1$ ns, i.e. 5
times better than at the raw data level.

\subsection{The off-line muon trigger}

After the PMT off-line calibration, thanks to the better time
accuracy of PMT hits, the SC trigger were re-applied to reject the
false coincidences found by the less accurate
on-line trigger algorithm.
Besides the SC trigger seed, other trigger seeds were searched in
each candidate muon event: the Floor Coincidence (FC) seed, that
is a coincidence between 2 hits recorded at the opposite ends of
the same storey ($\Delta T_{FC} \le 200$ ns) and the Charge
Shooting (CS) seed, that is a hit exceeding a charge threshold of
2.5 p.e..

The PMT hits of the event which satisfied at least one trigger seed
were, then, selected. Among this sub-sample, the number of
space-time causality relations ($N_{Caus}$) between ``trigger
hits" was calculated, using the following criterion:

\begin{equation}
|dt|<dr/v + 20 \hbox{ ns}, \label{eq:causal}
\end{equation}

where $|dt|$ is the absolute value of the time delay between two
hits, $dr$ is the distance between the two corresponding PMTs
(calculated using the Acoustic Positioning System data) and
$v$ is the group velocity of light in seawater.
The value of $N_{Caus}$ was used to reject events
containing only background hits. Only the events having
$N_{Caus}\ge4$ were considered for further analysis, according to
results of Monte Carlo simulations (see sec. \ref{sec:results}).
For the events passing this selection procedure, all PMTs hits
were re-considered in the following steps of the analysis.

\begin{figure}[htb]
\begin{center}
\includegraphics[width =7cm]{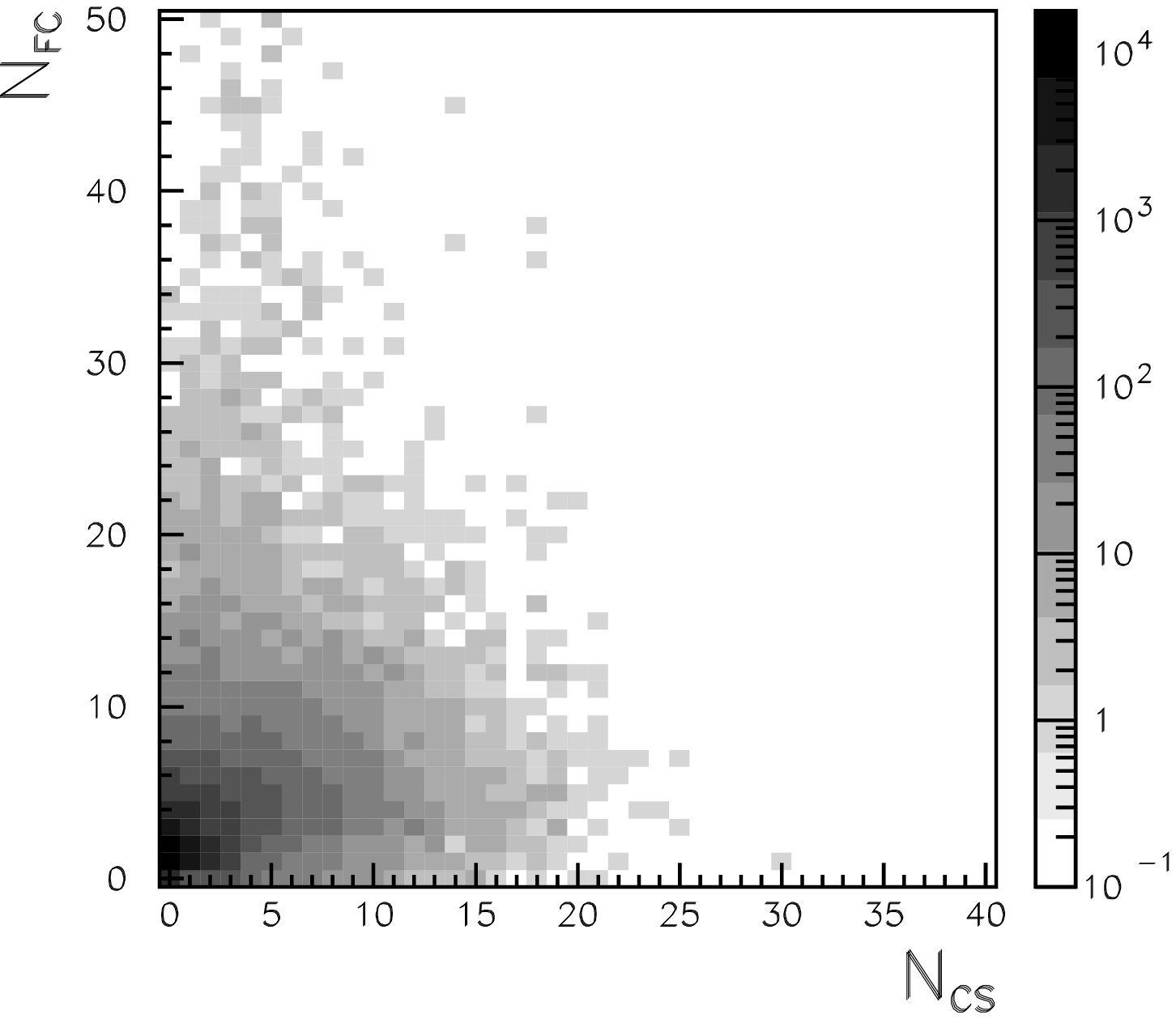}\includegraphics[width =7cm]{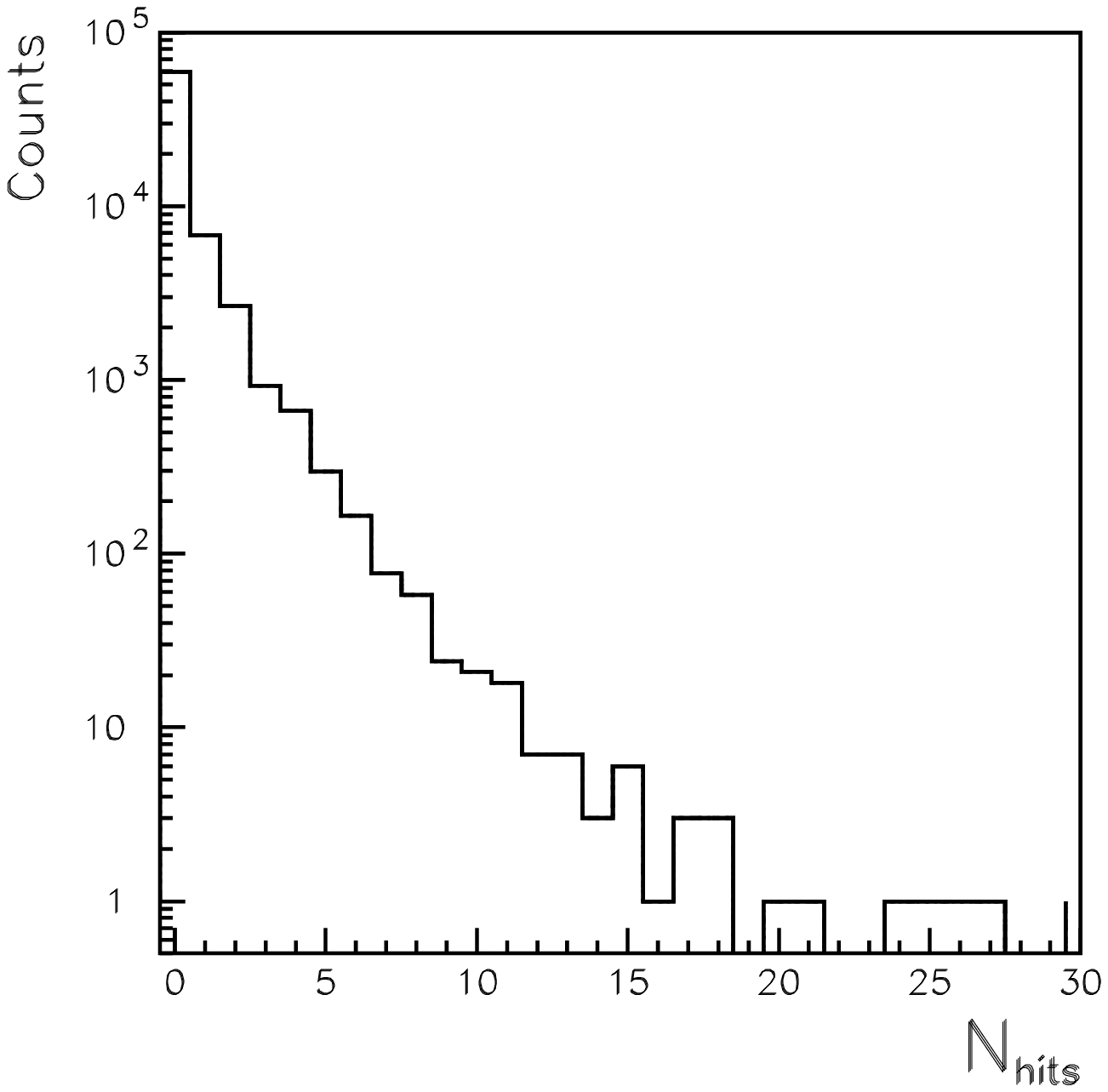}
\end{center}
\caption{Left: Contour plot showing the number of events as a
function of FC and CS trigger per event. Right: Number of hits
participating both in an FC and in a CS trigger (see text).}
\label{fig:SelectedeventsFCSC}
\end{figure}

In Fig. \ref{fig:SelectedeventsFCSC} is shown the correlation
between the number of FC and CS triggers for each event passing the
selection procedure. In the left-hand side the number of selected
events is plotted as a function of number of CS and FC triggers
per event. In the right-hand side, the number of hits, in the
whole data sample, participating in both an FC and in a CS trigger
is shown. A large fraction of the selected hits, do not satisfy
the two trigger conditions simultaneously.

\subsection{Muon track reconstruction}
\label{sec:causality}

\begin{figure}[htb]
\begin{center}
\includegraphics[width =7cm]{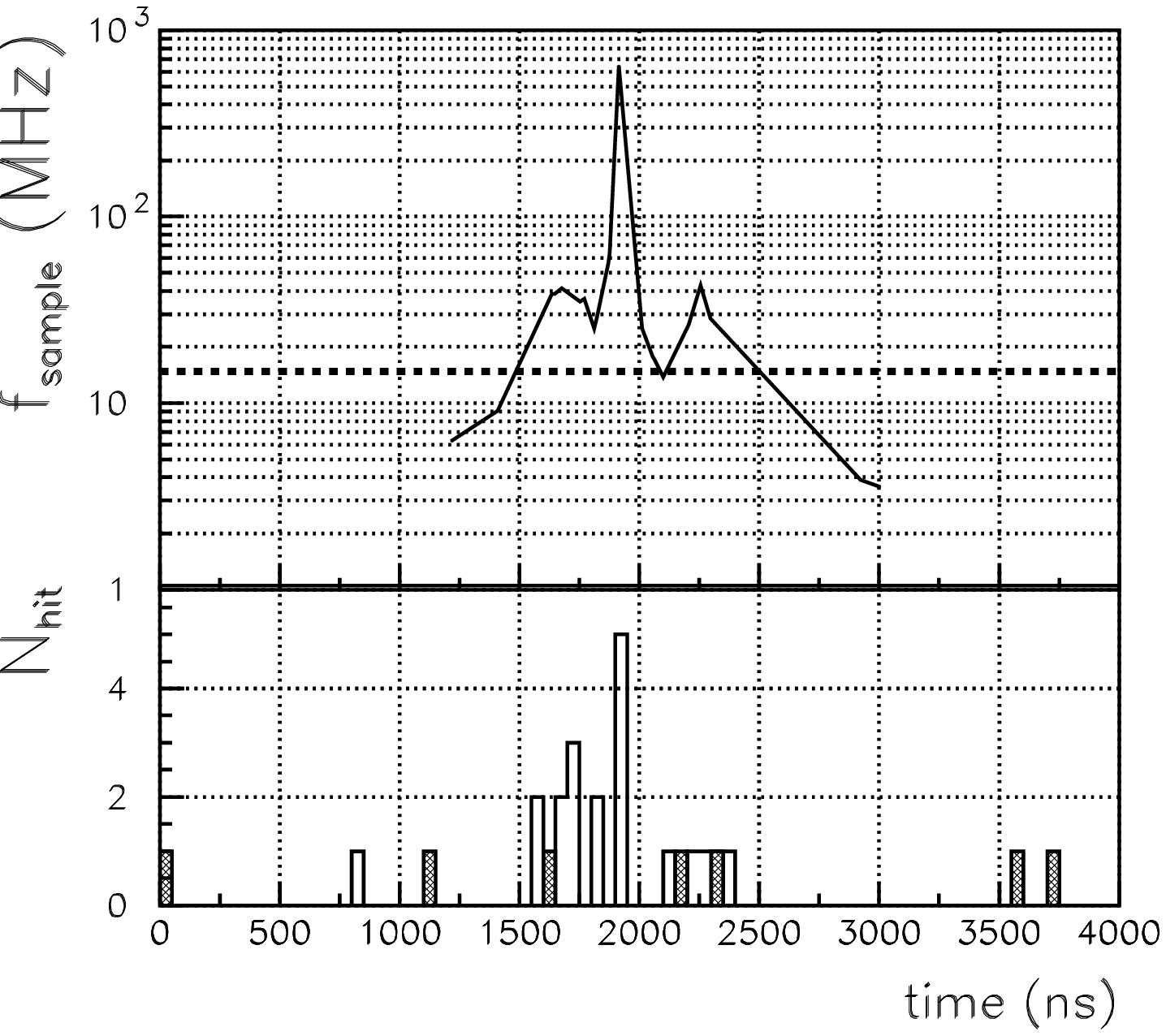}\includegraphics[width =7cm]{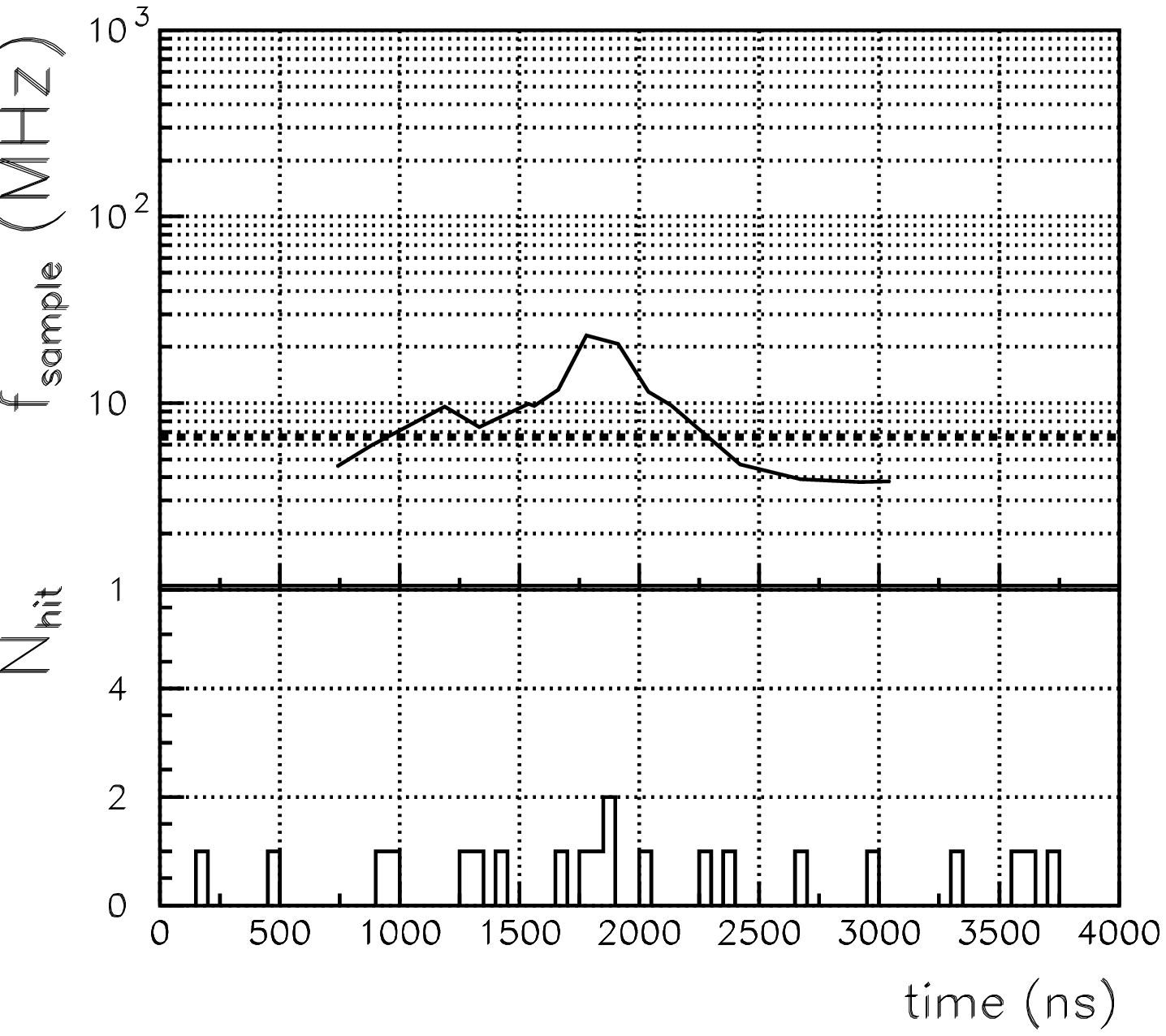}
\end{center}
\caption{Sample hit rate for $n=5$ as a
function of time for a simulated muon event (left) and for an
event in the data set (right). The dashed line represents the event average hit rate. The histograms in the left-hand
figure represent the distribution of muon hits (white filled) and
background hits (black filled). In the right-hand figure the histogram
represents real PMT hits (white filled).} \label{fig:hitsel}
\end{figure}

Once a muon event was identified, it was necessary to reject the
hits due to the optical background before starting the muon track
reconstruction procedure.
First, PMT hits with amplitude smaller than 0.5 p.e. were
rejected. Then, a new background filter algorithm \cite{Galata08},
was applied, as described in the following.
The $N$ PMT hits of the selected event, were sorted by arrival
time, and the {\it event average hit rate} was calculated as

\begin{equation}
f=\frac{N}{T_N-T_1},
\end{equation}

where $T_1$ and $T_N$ are the occurrence time of the ``earliest"
and the ``latest" hits in the selected event, respectively. PMT
hits were, then, grouped in samples of $n=5$ consecutive hits
(hit$_1$:hit$_5$, hit$_2$:hit$_6$, ..., hit$_{N-5}$:hit$_N$).
The hit rate of each sample $f_{sample}=n/\Delta t$, where $\Delta t=T_{k+n}-T_k$,
was calculated. The sample showing the highest hit rate 
%Since the total number of hits in the event is dominated by
%background, the Poisson expectation value of background hits for
%each sample is  given by $n_{exp}=f \cdot (T_{k+5}-T_k)$, where
%$T_k:T_{k+5}$ is the time interval of the sample (the index $k$
%running from 1 to $N-5$). At this point, the Poisson probability
%$P_{sample}$ to detect $n$ background hits was calculated for each
%sample. The sample with the minimum $P_{sample}$ 
is most likely to
contain muon hits, and it was taken as the reference sample for
the following steps of the muon reconstruction procedure. The
efficiency of this algorithm was proven via Monte Carlo
simulations (see sec. \ref{sec:results}). As an example Fig.
\ref{fig:hitsel} shows the distribution of hit arrival times for a
simulated (left-hand side) and for a real (right-hand side) muon
event crossing the Mini-Tower detector. For the simulated event,
the white filled histogram represents the muon-induced
\v{C}erenkov photons hits, while the black filled histogram refers
to optical background hits. For the real event PMT hits are shown
as a white-filled histogram. 
In both cases the upper plot shows the sample hit rate $f_{sample}$ with respect to the average event hit rate $f$. 
%the calculated Poisson probability $P_{sample}$.

The following step was to apply the causality relation in Eq.
\ref{eq:causal} between each of the $n$ hits of the reference
sample, and the other $N-1$ PMT hits in the event. A reference
hit, causally correlated with the largest number of the other
event hits, was chosen, and the ensemble formed by all the
causally correlated hits was stored and used to reconstruct the
atmospheric muon track. If this procedure preserved 6 hits, at
least, the muon track direction reconstruction codes were run.

The muon direction reconstruction algorithm used in this work  was originally developed by the ANTARES Collaboration and was then adapated to the
NEMO tower configuration  \cite{Heijboer03,Distefano07}. It is a
track fitting procedure based on a maximum likelihood method, that
takes into account the \v{C}erenkov light features and the
possible presence of non-rejected background hits.
The reconstruction strategy was applied in the way described below.

After the previous analysis cuts, hits participating to the trigger seeds could have been rejected, therefore 
the actual number of hits participating to the CS and CS trigger seeds was recalculated. 
These hits form the new ensemble on which a linear track pre-fit was applied.
At least three hits were required to compute the pre-fit. The results of the pre-fit were then used
as starting condition for the following fit algorithms (based on maximum likelihood method), that use
all the hits in the stored sample, and surviving the optical background filter.
During the likelihood fitting procedure other hits can be rejected and,
since the minimum number of hits required for track fitting is $N_{hit}\geq 6$, this reduces the total number of
reconstructed events\footnote{It is worth to mention that the applied likelihood reconstruction procedure is optimised for the larger size ANTARES  detector}.
Among all the tracks reconstructed with the likelihood method algorithm, only the ones satisfying the likelihood ($L$) condition:

\begin{equation}
 \Lambda\equiv\frac{\log_{10}(L) }{N_{\hbox{DOF}}} >-10
\end{equation} 

were selected, according to results of Monte Carlo
simulations, described in the next section.

\section{Results}
\label{sec:results}

In Tab. \ref{tab:ratevcut_data} the results of the muon
reconstruction procedure, at different stages of the analysis and
of the reconstruction procedure, are given. The table refers to
the sample of data recorded on 23-24 January 2007, shown in Fig.
\ref{fig:ratevstime}, when the tower was completely unfurled. 
\begin{table}[h]
\renewcommand{\arraystretch}{1.2}
\begin{center}
\caption{{\bf Results of the data analysis and track
reconstruction procedure:} number of events and reconstructed
tracks for each step of the analysis applied in sequence (see text). %The on-line trigger starts after a simple coincidence (SC). 
%The off-line trigger selects only events with $N_{Caus}\ge4$.  
} \label{tab:ratevcut_data} ~\\*[0.2cm]
\begin{tabular}{llc}
\hline
{\bf  LiveTime}                   				&					& $11.31$ hours\\
{\bf  On-line Trigger}            			&	at least 1 SC		& $6\cdot 10^7$ events \\
{\bf  Off-line Trigger}           			&	$N_{Caus}\ge4$	& $465386$ events\\
{\bf  Background Filter }  				&	$N_{hit}\geq$6		& $70913$ events\\
{\bf  Prefit reconstructed}                  		&	$N_{hit}\geq $3  ($hit  \in$ SC or CS)		& $13205$ tracks\\
{\bf  Likelihood reconstructed}              	&					& $3049$ tracks \\
{\bf  Selected }       					&	$\Lambda >-10$	& $1139$ tracks \\
\hline
\end{tabular}
\\*[1.cm]
\end{center}
\end{table}
The cuts applied on the data, at the level of the background filter
and likelihood fit parameter, where chosen according to the
results of the Monte Carlo simulation described below, optimized
for the data taken in the period 23-24 January 2007.

After the analysis cuts a total of 3049 atmospheric muon events
was reconstructed with an average muon
reconstruction rate of 0.075 Hz. 
The limited detector size and the
short livetime did not allow the detection of up-going
atmospheric neutrino events, whose rate is about 10$^{-6}$ times
less than for the muons. 

For the data recorded during the period 2 March - 12 April 2007
the acquisition livetime was 174.1 hours and the total number of
reconstructed atmospheric muons was 27699, thus the average
reconstruction rate was 0.044 Hz \cite{Cris08}. The lower rate of
reconstructed tracks was due to the smaller number of PMTs
participating in the muon reconstruction procedure, caused by the
improper tower configuration (see sec. \ref{sec:operations}).
These data were not used for further analysis.

\subsection{Detector Monte Carlo simulation}

In order to evaluate the detector response to atmospheric muons
and compare it with the results of the data analysis, a Monte
Carlo simulation was performed. Atmospheric muon events
were simulated using the MUPAGE code \cite{Carminati08}, a muon
event generator based on parametric formulas \cite{MupageFormula}.
Atmospheric muons were generated on the surface of a can-shaped
volume of water of $194$ m height and $238$ m radius (containing
the detector) in an energy range from 20 GeV to 500 TeV, according
to the MUPAGE limits. This energy range was suitable due
to the expected detector threshold and limited size.
A total of $4\cdot 10^7$ atmospheric muons was simulated, corresponding
to the real acquisition detector livetime of 11.3 hours. The generated
muon events were propagated inside the detector, using the
simulation tools developed by the ANTARES Collaboration
\cite{Becherini05}. These codes simulate the emission and
propagation of \v{C}erenkov light induced by muons and their
secondary products (e.g. showers and $\delta$-rays), then record
photo-electron signals on PMTs. As mentioned before, the actual
run conditions of the detector were taken into account. The
detector geometry was simulated using the PMT positions
reconstructed by the acoustic positioning system data. The light
absorption length as a function of photon wavelength was
introduced, according to the one measured at the detector site
\cite{sito}. Once the muon PMT hits -produced by \v{C}erenkov
photons originated along the muon track- were simulated, the
spurious PMT hits -due to the underwater optical noise- were
introduced in the simulation. It was assumed that the optical
background produces uncorrelated s.p.e. signals in the PMTs. In
the present work, optical background hits were simulated, for each
PMT, according to the real counting rate spectrum measured during
the selected period of detector operation (e.g. see Fig.
\ref{fig:rate}). The Mini-Tower DAQ electronics and the on-line
trigger were also simulated.

\begin{figure}[htb]
\begin{center}
\includegraphics[width =11cm]{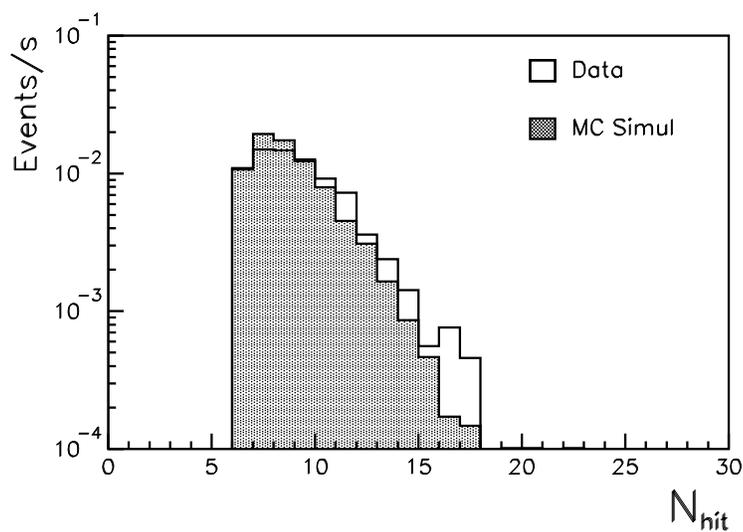}
\end{center}
\caption{Rate of muon tracks, as a function of the number of PMT hits, for events reconstructed from the data or from simulations, and Monte Carlo
events.} \label{fig:MC_data_comparison}
\end{figure}

Monte Carlo events surviving the on-line trigger simulation, were
processed using the same analysis chain used for the detector
data analysis. In Fig. \ref{fig:MC_data_comparison} the rate of
reconstructed tracks as a function of number of PMT hits, after
the causality filter, is shown both for Monte Carlo events and
detector data. Only the events preserving at least 6 PMT hits
could be processed by the track reconstruction algorithms.

\subsection{Comparison between data and simulations}

\begin{figure}[htb]
\begin{center}
\includegraphics[width =11cm]{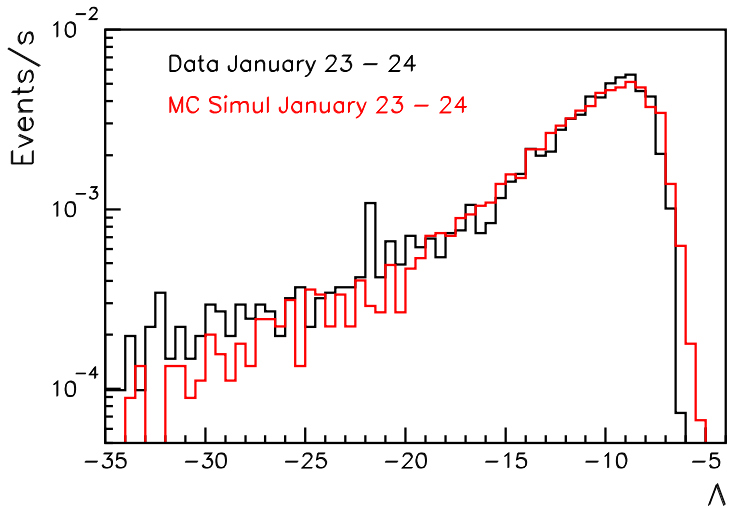}
\end{center}
\caption{Likelihood spectra of reconstructed muon tracks. }
\label{fig:lik}
\end{figure}

The good agreement found between data and Monte Carlo events is
shown in Fig. \ref{fig:lik}. The likelihood distribution of the
reconstructed muon track rate is shown both for data and Monte
Carlo events.

\begin{figure}[htb]
\begin{center}
\includegraphics[width =11cm]{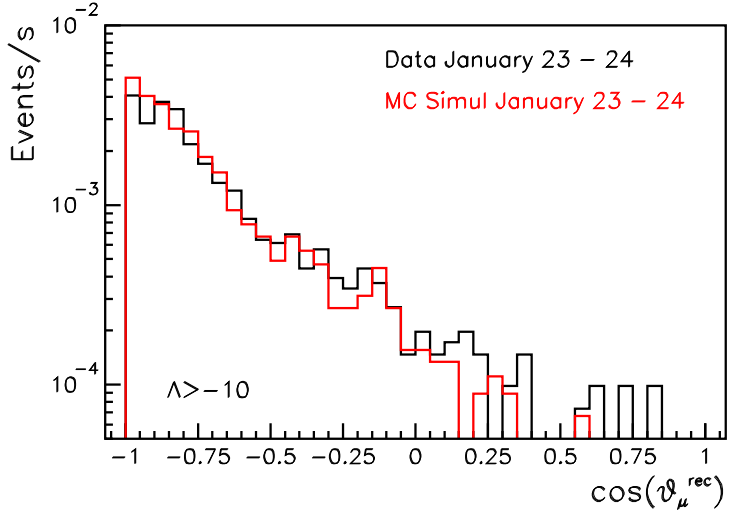}
\end{center}
\caption{Angular distributions of reconstructed muon track rate
after applying the likelihood quality cut $(\Lambda>-10)$.}
\label{fig:muatm}
\end{figure}

Among the reconstructed tracks, only the ones with a fit
quality parameter $\Lambda>-10$ were selected. In Fig.
\ref{fig:muatm} the zenith angular distribution of the
reconstructed muon track rate, after the fit quality cut, is
shown. The same distribution for Monte Carlo events is shown for
comparison. The Kolmogorov-Smirnov test \cite{Zech} was
performed\footnote{For this test the bin width was re-sized to
0.1.} to quantitatively evaluate the agreement between data and
Monte Carlo events, for the distributions plotted in Fig.
\ref{fig:muatm}. The test probability was found to be 0.81,
proving a good agreement between the two distributions.

\subsection{Depth Intensity Relation for Atmospheric Muons}

The final step of the analysis was the calculation of the so
called Depth Intensity Relation for the reconstructed atmospheric
muon tracks, that is the measurement of the vertical muon flux intensity as
a function of the muon slant depth in water. An atmospheric muon,
reaching the detector located at a depth $D$ from a Zenith angle
$\theta_Z$, propagates through a water slant $h$
\begin{equation}
h=\frac{D}{\cos\theta_Z}.
\end{equation}
The Depth Intensity Relation, thus, gives an estimate of the
vertical muon intensity as a function of the equivalent water
depth $h$ \cite{Amanda99}.

The muon intensity
$I(\theta_{\mu})$ as a function of the muon direction
$\theta_\mu$, was calculated using the relation
\begin{equation}
\label{eq:angflux}
I(\theta_{\mu})=  \frac{N_{\mu}(\theta_\mu) \cdot m(\theta_{\mu})} {{A_{eff}}(\theta_{\mu}) \cdot T \cdot \Delta \Omega}
\end{equation}
where
\begin{itemize}

\item  $N_{\mu}(\theta_\mu$) is the number of muon events assigned
by the analysis to the angular interval centered around $\cos
\theta_{\mu}$. For this, we considered the angular distribution
$N_{\mu}(\theta_\mu^{rec})$ obtained from the reconstruction,
without applying cuts. This distribution appears smeared because
of the detector angular resolution. We applied  an iterative
unfolding method based on Bayes' theorem \cite{Dagostini};

\item  $m(\theta_{\mu})$ is the mean muon multiplicity at the
angle $\theta_{\mu}$ and at the detector depth. It was determined
from the Monte Carlo simulations and it is shown in Fig.
\ref{fig:mult} ;

\begin{figure}[h]
\begin{center}
\includegraphics[width =9.cm]{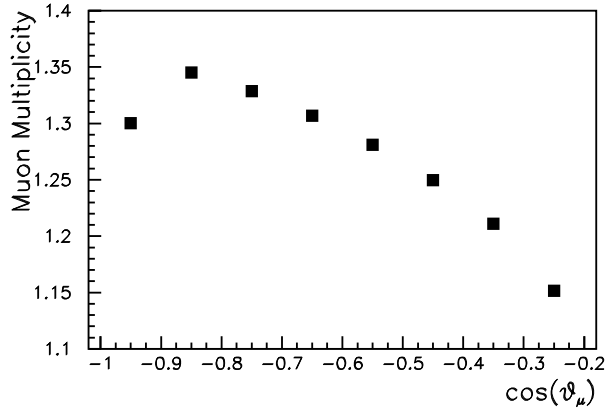}
\end{center}
\caption{Mean muon multiplicity as a function of the 
angle $\theta_{\mu}$ and at the detector depth.}
\label{fig:mult}
\end{figure}

\item $A_{eff}(\theta_{\mu})$ is the reconstruction effective area
at the muon angle $\theta_{\mu}$, determined by Monte Carlo
simulations and shown in Fig. \ref{fig:aeff};

\begin{figure}[h]
\begin{center}
\includegraphics[width =9.cm]{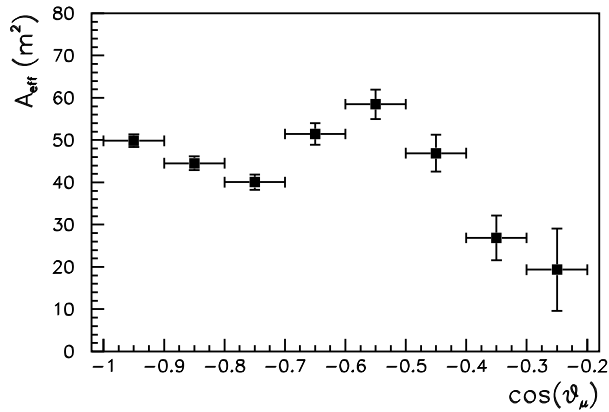}
\end{center}
\caption{Reconstruction effective area as a function of
the muon angle $\theta_{\mu}$, determined by Monte Carlo
simulations.} \label{fig:aeff}
\end{figure}

\item  $T$ is the data livetime. We used only the data acquired in
the period on 23-24 January 2007 with $T=11.3$ hours;

\item  $\Delta\Omega$ is the solid angle covered by the
corresponding $\cos{\theta_{\mu}}$ interval.

\end{itemize}

\begin{figure}[h]
\begin{center}
\includegraphics[width =9.cm]{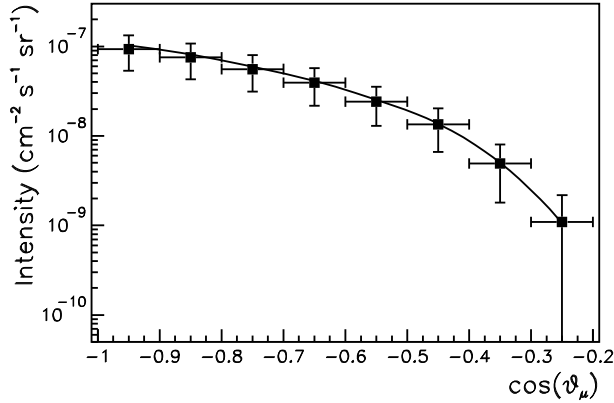}
\end{center}
\caption{Angular distribution of the atmospheric muon flux,
$I(\theta_{\mu})$, computed with Eq. \ref{eq:angflux}, for the
data acquired in the period 23-24 January 2007. The errors include
statistical and systematic uncertainties added in quadrature. Data
are compared with the simulated atmospheric muon flux (solid
line). } \label{fig:angdir}
\end{figure}

Fig. \ref{fig:angdir} shows the angular distribution of the
atmospheric muon flux obtained from Eq. \ref{eq:angflux}. Error
bars include both statistical and systematic uncertainties added
in quadrature. Systematic errors were evaluated via Monte Carlo
simulation, taking into account the uncertainties on the input
parameters: the light absorption length in water ($L_a$), the
light scattering length in water ($L_b$), the PMT quantum
efficiency and the angular acceptance of the Optical Module
\cite{Leonora09,antaresOM}. In Tab. \ref{tab:errors} the contributions of the different parameters to the total systematic error are reported.
The systematic error is mainly a scale error common to all measured points; its contribution to the point to point error is considerably smaller.

\begin{table}[h]
\renewcommand{\arraystretch}{1.2}
\begin{center}
\caption{{\bf  Systematic error:}  Contribution to the
systematic error due to the uncertainty on each input parameter of
the Monte Carlo simulation.} \label{tab:errors} ~\\*[0.2cm]
\begin{tabular}{lcc}
\hline
\hline
   {\bf  Input Parameter} & {\bf Relative Uncertainty of the Parameter} & {\bf $\Delta I/I$}  \\
\hline
$L_a$                        & $\pm 10$\%        &   $^{+20\%}_{-27\%}$  \\
$L_b$                        & $\pm 10$\%        &   $^{+3\%}_{-19\%}$   \\
PMT Quantum efficiency       & $\pm 10$\%        &   $^{+21\%}_{-15\%}$  \\
OM Angular Acceptance        & $\pm 10^\circ$    &   $^{+22\%}_{-8\%}$   \\
\hline
Total 	&	&	$^{+36\%}_{-37\%}$   \\
\hline
\end{tabular}
\\*[1.cm]
\end{center}
\end{table}

The measured flux $I(\theta_\mu)$ was, then, transformed into muon
vertical flux intensity $I(\theta_Z=0,h)$ using the formula:
\begin{equation}
I(\theta_Z=0,h)=I(\theta_Z) \cdot \cos(\theta_Z) \cdot c_{corr},
\end{equation}
where the zenith angle is $\theta_Z=180^\circ - \theta_\mu$. The
term  $c_{corr}$ is a correction factor required for angles larger
than 60$^\circ$ as described in  \cite{Lipari}.

\begin{figure}[h]
\begin{center}
\includegraphics[width =8.5cm]{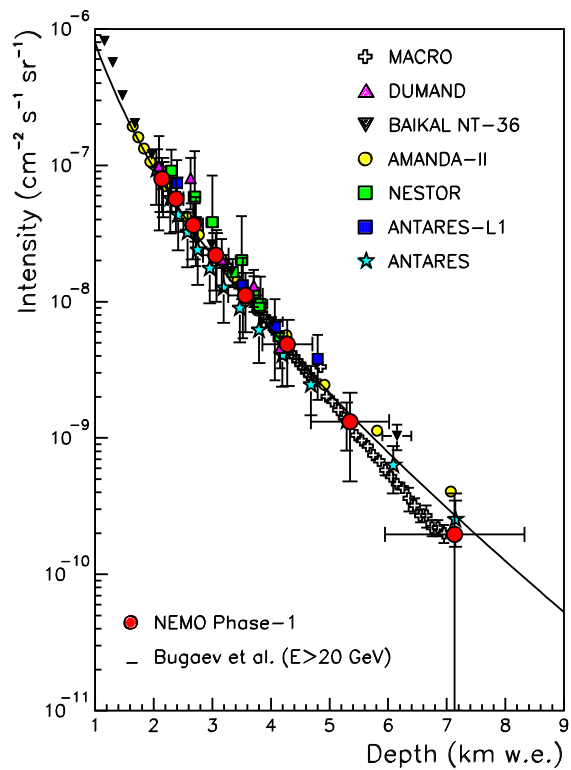}
\end{center}
\caption{Vertical muon intensity, $I(\theta_Z=0,h)$, versus depth
measured using data acquired in the period 23-24 January 2007. For
comparison, results from other experiments are quoted. The solid
line is the prediction of  Bugaev et al. \cite{Bugaev}.}
\label{fig:dir}
\end{figure}

In Fig. \ref{fig:dir} we show the Depth Intensity
Relation for atmospheric muons. Results obtained by previous experiments are shown for
comparison: MACRO \cite{Macro} in standard rock, DUMAND
\cite{Dumand}, NESTOR \cite{Nestor}, ANTARES \cite{AntaresL1,
AntaresDirIcrc}\footnote{ANTARES has also recently presented a DIR curve based on a novel data analysis method \cite{Zaborov}, whose results are not quoted in Fig. \ref{fig:dir}.}  in sea water, BAIKAL \cite{Baikal} in lake water,
AMANDA \cite{Amanda2} in ice. Results are also compared with the
prediction of Bugaev et al. \cite{Bugaev}. The NEMO Phase-1 data
are in agreement with previous measurements and with
Bugaev's prediction in the whole range of investigated depths.

\section{Conclusions}

The activities of the NEMO Collaboration have progressed with the
achievement of major milestones: the realization and installation
of the Phase-1 apparatus.

With this apparatus it was possible to test in deep sea the main
technological solutions developed by the collaboration for the
km$^3$ scale underwater neutrino telescope.

The angular distribution of atmospheric muons was measured and
results were compared to Monte Carlo simulations. The vertical
muon intensity was evaluated and compared with previous data and
predictions, showing a good agreement.

\section*{Acknowledgments}

We thank all INFN personnel that has contributed to the
development and carry out mechanics, electronics and computing of
the NEMO Phase-1 experiment. We also thank the ANTARES
Collaboration for providing the detector simulation and track
reconstruction codes, extensively used in this work. 
Eventually we thank our anonymous referee for useful comments and suggestions.

\end{document}